# Competition between delamination and fracture in multiple peeling problems

Lucas Brely[(1)], Federico Bosia[(1)], Ali Dhinojwala[(2)] and Nicola M. Pugno[(3,4,5)*]

*(1)  Department of Physics and "Nanostructured Interfaces and Surfaces" Inter-Departmental Centre, Università di Torino, Via P. Giuria 1, 10125, Torino (Italy).*

*(2)  Department of Polymer Science, The University of Akron, Akron, Ohio 44325-3909*

*(3)  Laboratory of Bio-Inspired & Graphene Nanomechanics, Department of Civil, Environmental and Mechanical Engineering, Università di Trento, via Mesiano, 77, I-38123 Trento, Italy.*

*(4)  School of Engineering and Materials Science, Queen Mary University of London, Mile End Road, London E1 4NS.*

*(5)  KET Labs, Edoardo Amaldi Foundation, Via del Politecnico snc, 00133 Rome, Italy*

(\*) Corresponding author: *nicola.pugno@unitn.it*

**Keywords:** Adhesion, fracture, multiple peeling, Simulations, Spider web

**Abstract**

Adhesive attachment systems consisting of multiple tapes or strands are commonly found in nature, for example in spider web anchorages or in mussel byssal threads, and their structure has been found to be ingeniously architected in order to optimize mechanical properties, in particular to maximize dissipated energy before full detachment. These properties emerge from the complex interplay between mechanical and geometric parameters, including tape stiffness, adhesive energy, attached and detached lengths and peeling angles, which determine the occurrence of three main mechanisms: elastic deformation, interface delamination and tape fracture. In this paper, we introduce a formalism to evaluate the mechanical performance of multiple tape attachments in different parameter ranges, allowing to predict the corresponding detachment behaviour. We also introduce a numerical model to simulate the complex multiple peeling behaviour of complex structures, illustrating its predictions in the case of the staple-pin architecture. We expect the presented formalism and numerical model to provide important tools for the design of bioinspired adhesive systems with tunable or optimized detachment properties.

1. **Introduction**

Spider silk is a biological fibrous material that displays exceptional mechanical properties [1] and comes in many different types, each with specific functions and properties [2]. Silk is the base construction material of spiders, and is used to fabricate complex structures such as the spider web. In addition to the main structure, the attachment between the silk threads and the substrate plays an important role in determining the functionality of silk-based architectures. For example, it was shown that the contact, usually performed through adhesive-coated "silken" threads called "attachment discs" [3] differs in geometrical features depending on its prey-capture or locomotion functions [4]. To create a safe attachment between the dragline and the substrate, spiders create a structure referred to as a "staple-pin" attachment. An array of perpendicular silken threads are used to "coat" the main thread on which the external load is applied (see Fig. 1A) [5]. When staple-pin structures are subjected to a peel test, different types of behaviour have been observed in natural systems [6]. The detachment occurs in some cases by delamination of the secondary tapes, which corresponds to the failure of the interface between the system and the substrate, and in other cases by the breakage of the secondary tapes themselves. The occurrence of these two mechanisms suggests that the adhesive energy of the tape/substrate contact is high with respect to the fracture energy of the adhesive tapes, and that elastic deformation plays an important role in the total dissipated energy under load of the staple-pin system. The compliance of the adhesive tapes, associated with a low contact angle in such structures has been attributed to a spider strategy to develop maximum adhesive strength out of a minimum amount of material and artificial systems mimicking the spider attachment disc have recently been introduced [7], with the aim of optimizing the maximum detachment force and the total dissipated energy out of minimal contact area and material use.

Various theoretical approaches have been developed to treat thin film-peeling problems in the case of single [8] [9] and multiple tapes [10]. The objective of the latter is to describe the behaviour of a system containing various simultaneously detaching tapes [11], which apply to natural systems like gecko toes [12] as well as spider we anchorages. However, the behaviour under loading of multiple peeling systems is not trivial and ad hoc numerical procedures are required to simulate their delamination [13] [14]. Various numerical approaches have been developed to address specific problems, such as adhesion to various types of surfaces [15], influence of hierarchical structure [16], role of friction [17] and viscoelasticity [18]. These numerical modelling tools are essential to design bioinspired artificial micro-patterned surfaces with optimized properties [19], including hierarchical structures [20].

Here, we develop analytical and numerical models to simulate the delamination and failure of coupled adhesive tapes, also taking into consideration the elastic deformation and the peeling angle variation under load of the staple-pin attachment. We propose a general numerical scheme to model the detachment of staple-pin-like structures, introducing new aspects to existing models such as tape fracture and 3-D deformation of the attachment devices. This work can help develop new designs for efficient bio-inspired attachments, maximizing adhesive strength while minimizing material use.

## 2. Model

We consider the geometry shown in *Figure 1*. The attachment system is built from an array of tapes attached perpendicularly to the main cable on which a vertical external load is applied. Considering a single tape from the staple-pin structure, the problem reduces to studying the symmetric double peeling system shown in Figure 1B [21].

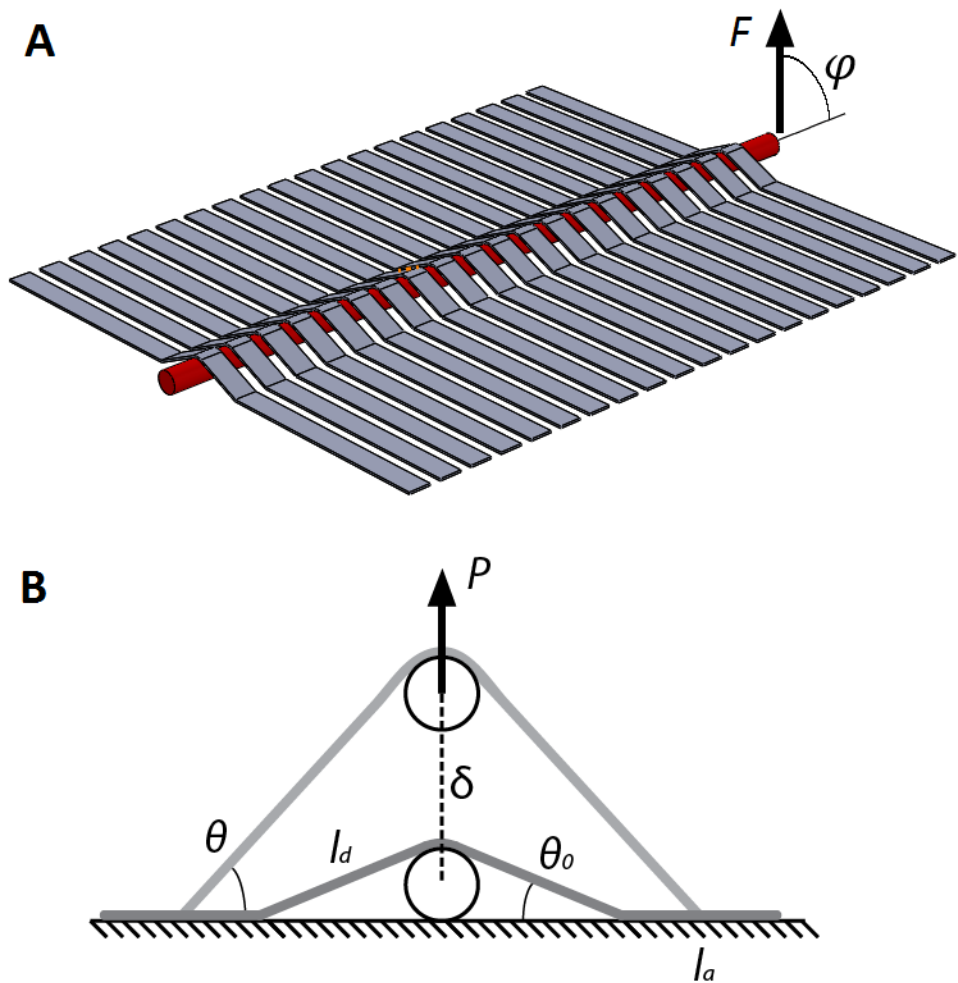

*Figure 1: A. The staple-pin attachment structure: F is the applied external load, $\varphi$ its angle with respect to the plane. B. The symmetric double peeling system viewed as a sub-domain of the staple-*

pin. *P is the applied load, $l_a$ is the attached tape length, $l_d$ is the detached tape length, $\delta$ is the vertical displacement of the load application point, $\theta_0$ is the initial peeling angle, $\theta$ is the current peeling angle.*

The detachment of tapes adhering to a substrate can be theoretically described by using energy balance considerations, i.e. the peeling front advances when the detachment leads to a decrease of the total potential energy $\Pi_p$ in the system:

$$\frac{\partial \Pi_P}{\partial l_d} < 0 \tag{1}$$

where $l_d$ is the detached length of the tape. Introducing the surface energy $U_s$, the stored strain energy in the tape $U_e$ and the work associated with the external load acting on the system $V$, Eq. (1) becomes:

$$\frac{\partial V}{\partial l_d} - \frac{\partial U_e}{\partial l_d} > \frac{\partial U_s}{\partial l_d} \tag{2}$$

We first consider an inextensible film, in which the contribution of the stored elastic energy is neglected ($U_e = 0$), which leads to the solution of the Rivlin equation [8]. According to the latter, delamination occurs when:

$$T(1 - \cos\theta) = wG \tag{3}$$

where $T$ is the load applied to the tape, $\theta$ is the peeling angle, i.e. the angle between the tape and the substrate, $w$ the tape width and $G$ the adhesive energy per unit length of the contact. For the symmetric V-shaped double peeling system in Fig.1.B, the applied external load $P$ is:

$$P = 2\sin\theta\, T \tag{4}$$

and peeling starts when the external load reaches the value $P_1$

$$P_1(\theta) = \frac{2wG \sin \theta}{1 - \cos \theta} \tag{5}$$

Overall, from an initial tape-substrate angle configuration $\theta_0$, the external load is applied and the peeling angle changes as the structure detaches. This behaviour is shown in a $P$ vs. $\theta$ plot in *Figure 2.A*, for $w = 10$ mm and $G = 0.1$ MPa·mm. Starting from an unloaded structure ($P(\theta_0 = \pi/2) = 0$, point O) and increasing $P$, the tapes will start to peel off at $P = P_1$ (point A), leading to a decrease in the peeling angle and consequently an increase in $P_1$. In this case, the peeling angle tends to zero as the delamination proceeds and the peeling force increases indefinitely. The admissible space of load-angle configurations is $\Omega = \{P \leq P_1\}$, shown as the shaded area in *Figure 2.A*.

We now introduce a critical tape tension $T_c$ at which the tape fractures. Depending on the adhesive energy and the geometrical and mechanical properties of the system, three different behaviours can be observed as a function of the adhesive energy $G$ (*Figure 2B*). When $G$ is smaller than a given value $G_1$ ($G < G_1$), the tape delaminates over its full attached length $l_a$ and the critical tension is not reached during the process. From Eq.(3):

$$G_1 = \frac{T_c(1 - \cos \theta)}{w} = \frac{T_c}{w}\left(1 - \frac{l_d \cos \theta_0 + l_a}{l_a + l_d}\right) \tag{6}$$

If $G$ is greater than a second given value $G_2$ ($G > G_2$) the tape fractures before any delamination (and thus angle change) occurs. This happens when the external load reaches a critical value $P_c = 2 \sin \theta T_c$ (also shown in *Figure 2A*). Again, from Eq.(3) with $\theta = \theta_0$, we have:

$$G_2 = \frac{T_c(1 - \cos \theta_0)}{w} \tag{7}$$

If $G$ lies between $G_1$ and $G_2$, i.e. $G_1 < G < G_2$, the critical tension is reached after a finite delamination length.

The dissipated energy $W$ due to delamination depends on the adhesive energy and the corresponding tape delamination behaviour:

$$W = \int_\eta P \, d\eta \tag{8}$$

where $\eta$ is the displacement of the external load application point. Here, we neglect the dissipated energy due to tape fracture. When $G > G_2$, the dissipated energy associated with the fracture of the tapes with no delamination is zero, i.e. $W_2 = 0$. On the other hand, full delamination with no fracture is obtained when $G < G_1$ and the dissipated energy due to delamination is:

$$W_{G<G_1} = 2l_a w G \tag{9}$$

If the adhesive energy reaches the value $G = G_1$, the entire attached region peels off and the tape fractures at the final delamination point. This corresponds to the maximum of dissipated energy:

$$W_{G_1} = 2l_a w G_1 \tag{10}$$

For intermediate $G$ values between $G_1$ and $G_2$, the tape delaminates for part of its attached length and breaks when the applied load reaches $P_c$, with the dissipated energy $W$ decreasing linearly with delaminated tape length. The three cases are illustrated in *Figure 2B*, where the dissipated energy is plotted as a function of the adhesive energy using Eq. (6) and Eq. (7) adopting the test parameters $l_{d0} = 1$ mm, $l_a = 50$ mm and $T_c = 2$ N. The limit values $G_1$ and $G_2$ are both highlighted.

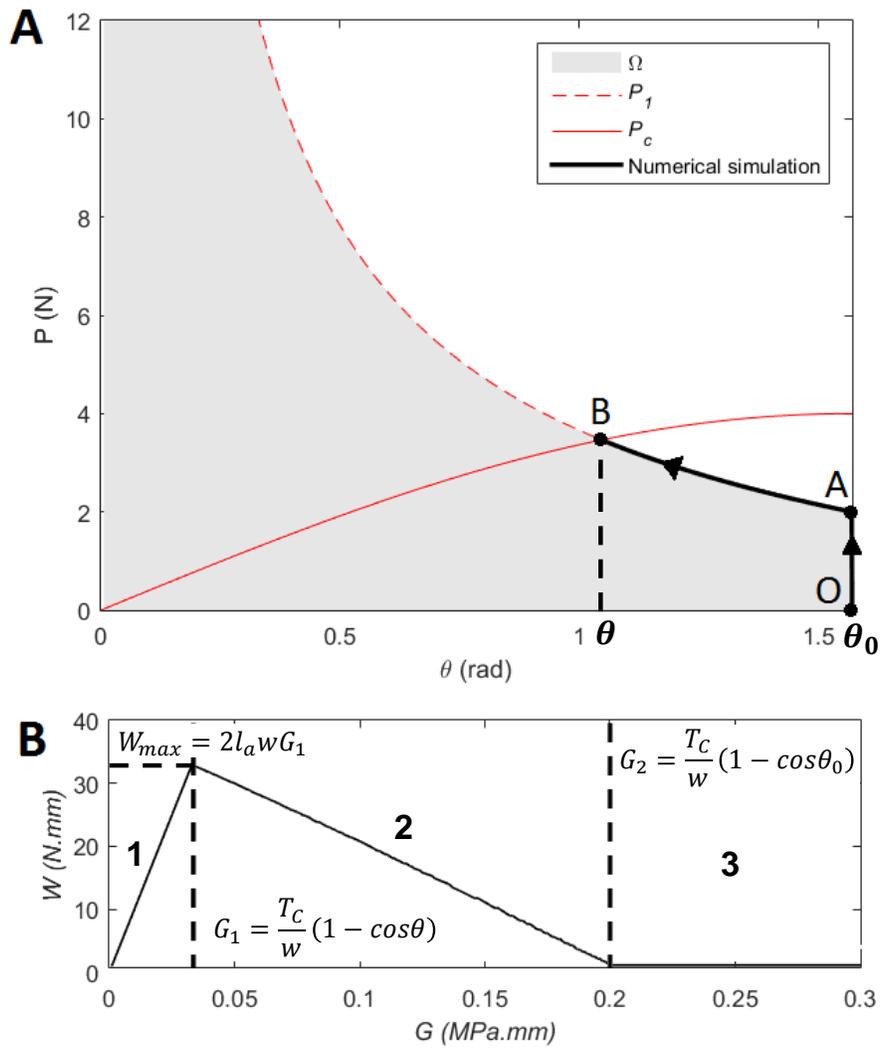

Figure 2: A. Rigid tape symmetric double peeling behaviour in the peeling angle / external load space. Shaded area represents admissible values. $P_1$ is the delamination load, $P_c$ the critical (fracture) load. B. Dissipated energy W as a function of the adhesive energy G. Three behaviours can be distinguished: 1- Delamination ($G<G_1$), 2- Delamination and fracture ($G_1<G<G_2$), 3- Fracture without delamination ($G>G_2$).

Considering now additionally tape elastic deformation and its contribution to energy balance, Eq. (1) now becomes the Kendall equation [9]:

$$T(1 - \cos\theta) + \frac{T^2}{2Ewb} = wG \qquad (11)$$

where $b$ is the tape thickness and $E$ is the tape elastic modulus. In this case, the peeling angle changes as a function of the elastic deformation and the detachment of the structure. From a given initial configuration, it is possible to write the relationship between the tape strain $\varepsilon$ and the tape-substrate angle $\theta$ as:

$$l_d(1 + \varepsilon)\cos\theta = l_d \cos\theta_0 \qquad (12)$$

Writing the tape tension as $T = Ewb\varepsilon$, we obtain:

$$T = Ewb\left(\frac{\cos\theta_0}{\cos\theta} - 1\right) \qquad (13)$$

The external load $P$ for the double peeling structure in Fig.1B is:

$$P(\theta) = 2\sin\theta\, T = 2\sin\theta Ewb\left(\frac{\cos\theta_0}{\cos\theta} - 1\right) \qquad (14)$$

which assumes its maximal value $P_2$ for an initial peeling angle of $\theta_0 = 0$. The overall delamination problem can be treated as the superposition of two independent single peeling processes, with each tape loaded by its peeling tension [10]. The external load at delamination $P_1$ for a given adhesive energy is:

$$P_1(\theta) = 2\sin\theta Ewb\left(\cos\theta - 1 + \sqrt{(1-\cos\theta)^2 + \frac{2G}{Eb}}\right) \qquad (15)$$

When $P \geq P_1$, delamination occurs. The maximum peeling load is therefore obtained for $\theta = 0$. From this, we obtain the admissible space of load-angle configurations as $\Omega = \{P(\theta) \leq P_2(\theta) \cap$

$P(\theta) \leq P_1(\theta)\}$, shown as the shaded area in *Figure 3.A*. For a given unloaded structure ($P(\theta_0) = 0$, point O), the system will first undergo elastic deformation without delamination until $P = P_1$ (point A), and then both elastic deformation and delamination which leads the system to the equilibrium state (point B). The limit angle $\theta_{lim}$ in the equilibrium state is obtained when $P_1 = P_2$, which using Eq. (14) for $\theta_0 = 0$ and Eq. (15), leads to the following relation:

$$2\cos^3\theta_{lim} - \left(3 + \frac{2G}{Eb}\right)\cos^2\theta_{lim} + 1 = 0 \tag{16}$$

Note that when $\theta_0 = 0$, the equilibrium state is reached as soon as the first delamination occurs. In this particular case, the stored elastic energy per unit length remains constant and no angle variation occurs during delamination. The results plotted as example in *Figure 3.A* are relative to the parameters $w = 0.01$ mm, $b = 10$ mm, $E = 100$ MPa.

The dissipated energy $W$ in the elastic tape case is:

$$W = U_{el} + U_{del} = \int_\eta P\, d\eta \tag{17}$$

Where $U_{el}$ is the elastic energy stored in the deformed tape when complete detachment or fracture occurs. Here, the problem becomes more complex than the rigid tape case, and the analytical description of the system behaviour must be simplified by assuming that the attached length is sufficiently long to reach the previously described equilibrium state. Under this assumption, the critical tape tension $T_C$ is first reached (i.e. full delamination and then fracture occur) when $G = G_1$ and:

$$T_C = Ewb\left(\cos\theta_{lim} - 1 + \sqrt{(1 - \cos\theta_{lim})^2 + \frac{2G_1}{Eb}}\right) \tag{18}$$

From Eq. (18), we obtain

$$\cos\theta_{lim} = \frac{T_C}{2Ewb} + 1 - \frac{G_1 w}{T_C} \qquad (19)$$

Combining Eq. (19) with Eq. (13) for $\theta_0 = 0$, it is possible to determine $G_1$ for $\theta = \theta_{lim}$:

$$G_1 = \frac{T_C^3 + 3EwbT_C^2}{2E^2b^2w^3 + 2T_C Ebw^2} \qquad (20)$$

The corresponding dissipated energy is the sum of the adhesive and elastic energy:

$$W_1 = 2l_a w G_1 + 2(l_a + l_d)\frac{T_C^2}{2Ewb} \qquad (21)$$

On the other hand, the adhesive energy $G_2$ beyond which fracture occurs without delamination can be obtained from Eq. (11) as:

$$G_2 = \frac{T_C^2}{2Ebw^2} + \frac{T_C}{w} \qquad (22)$$

for which the dissipated energy is:

$$W_2 = 2l_d \frac{T_C^2}{Ewb} \qquad (23)$$

Notice that these expressions for $G_1$ and $G_2$ tend to the values for the rigid tape case for $E \to \infty$. Numerically calculated curves for $\theta_0 \neq 0$ are included in *Figure 3B*. When $G < G_1$, $W$ is independent of the initial angle $\theta_0$. When $G > G_1$, the dissipated energy tends to decrease as the initial angle decreases. This is because the tape fracture load is reached sooner.

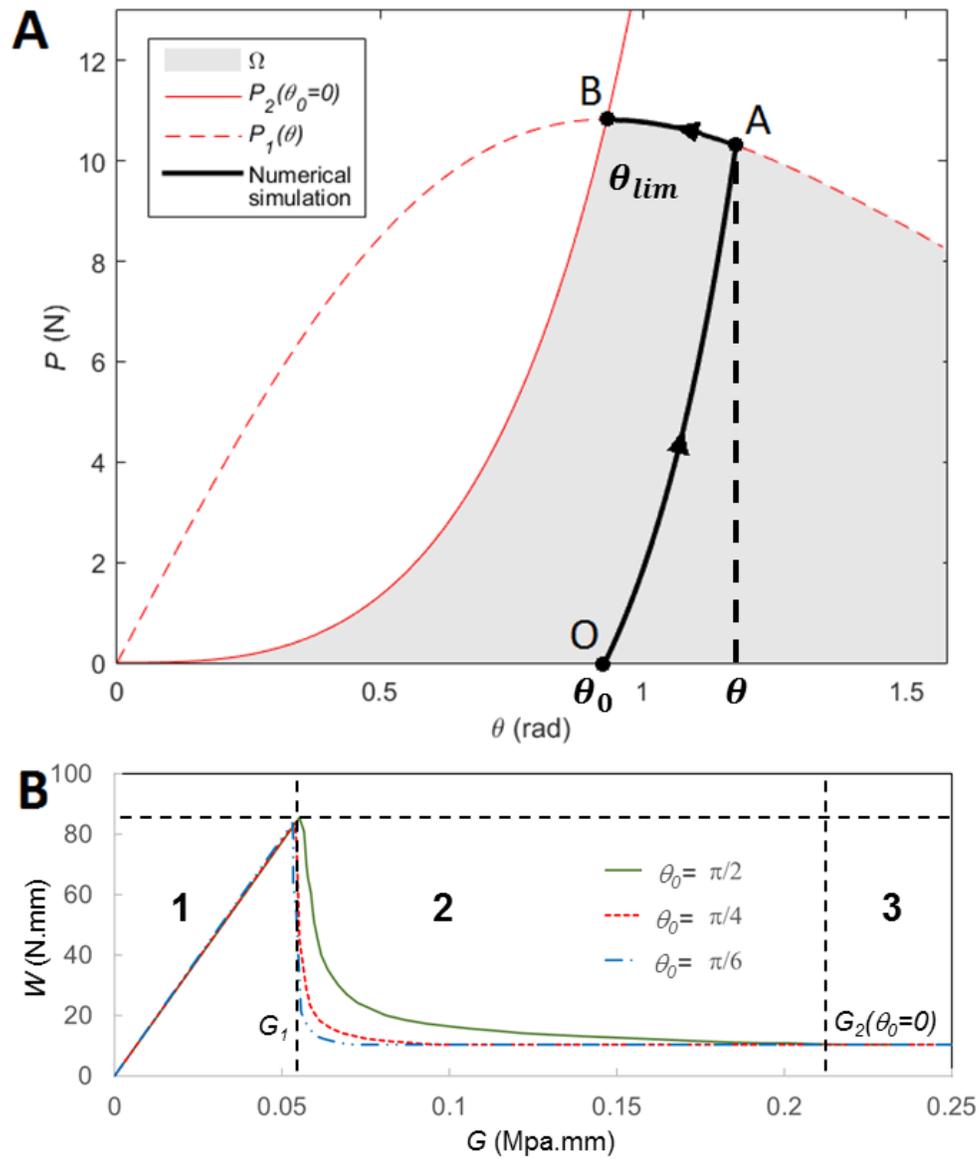

*Figure 3: Elastic tape symmetric double peeling behaviour: A. External load vs. peeling angle in the case $G = G_1$(tape fracture after full delamination); $P_1$ is the delamination load, $P_2$ the load under deformation for $\theta_0 = 0$. B. Dissipated energy by the system W as a function of the adhesive*

*energy G for various initial peeling angles $\theta_0$. As in Fig. 2, three behaviours can be distinguished: 1- Delamination ($G<G_1$), 2- Delamination and fracture ($G_1<G<G_2$), 3- Fracture ($G>G_2$).*

The load vs. displacement curve in the optimal case $G = G_1$ is shown in *Figure 4*, together with the contribution of the elastic and delamination energy. The resulting curve displays a perfect elastoplastic behaviour.

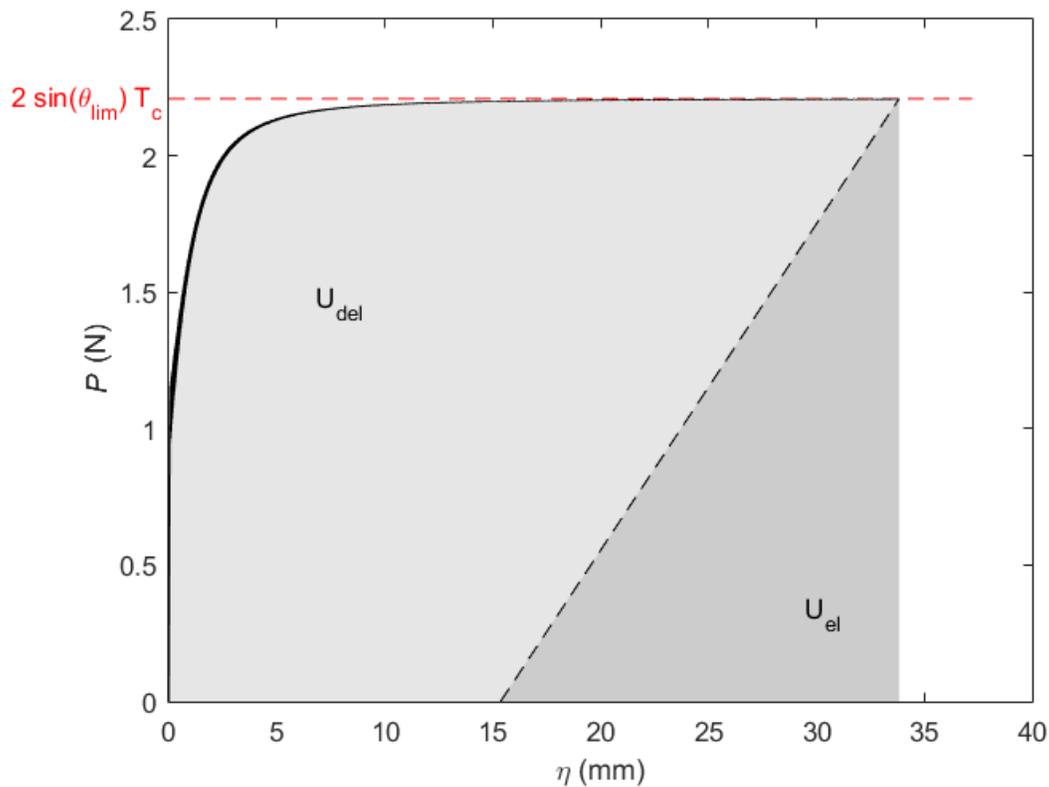

*Figure 4: External load as a function of the displacement at the load application point during the detachment of a symmetric double peeling elastic tape when the adhesive energy is optimal. The energy dissipated during detachment and the stored elastic energy are shown as shaded areas.*

Notice that the above discussion can be generalized from a 2D to 3D structure, in which the deformed detached length of the tape is not aligned with the delamination direction, as illustrated in *Figure 5*. In this case, Eq. (3) is modified as follows:

$$T(1 - \cos\theta \cos\lambda) = wG \qquad (24)$$

where $\lambda$ is the angle defining the misalignment of the detached tape and attached length, due to deformation or initial conditions. Since the elastic energy variation and the surface energy remain unchanged, Eq. (11) becomes:

$$T(1 - \cos\theta \cos\lambda) + \frac{T^2}{2Ewb} = wG \qquad (25)$$

When both attached and detached tape are aligned ($\lambda = 0$) the above equation coincides with Eq. (11). Solving Eq. (25) provides the tension needed to detach the tape:

$$T_1 = 2Ewb\left(\cos\theta \cos\lambda - 1 + \sqrt{(1 - \cos\theta \cos\lambda)^2 + \frac{2G}{Eb}}\right) \qquad (26)$$

These equations are valid under the hypothesis that even when $\lambda \neq 0$, the load is equally distributed over the peeling line, and that the width of the peeling line remains unchanged (with respect to $\lambda = 0$). This could be an over-simplification of the problem in some cases, and local load concentrations could appear, with a reduction of the peeling force.

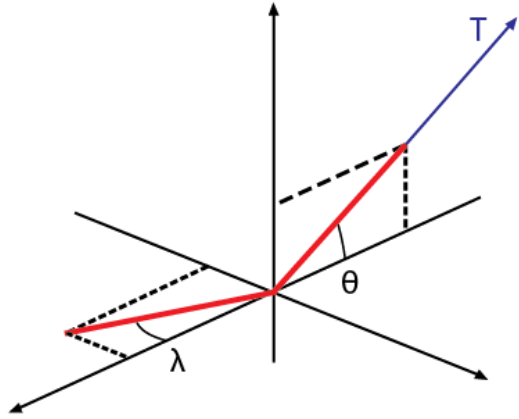

Figure 5: Misalignment between the attached and detached tape.

3. Numerical implementation

To simulate the delamination and fracture behaviour of arbitrary multiple tape structures, we adopt a general numerical model based on the mechanical equilibrium and energy balance. For a given structure in 3-D space, mechanical equilibrium is obtained using the co-rotational truss formulation [22]. The system is built using a frame of truss members sustaining axial load only, where the elements in contact with the substrate act as peeling tapes. A member $k$ of the system linking the nodes $i$ and $j$, is defined by its initial length $l$, thickness $b$, width $w$, elastic modulus $E$ and its direction cosines in 3-D space:

$$c_1 = \frac{x_j - x_i}{l_k} \tag{27}$$

$$c_2 = \frac{y_j - y_i}{l_k} \tag{28}$$

$$c_3 = \frac{z_j - z_i}{l_k} \tag{29}$$

Where $\mathbf{x} = [x \quad y \quad z]$ is the coordinate vector of a node under deformation of the system. The contribution of the member $k$ to the material and geometric stiffness matrix (respectively $\mathbf{K_m}$ and $\mathbf{K_g}$) is then obtained as:

$$\mathbf{L} = [c_1 \quad c_2 \quad c_3 \quad -c_1 \quad -c_2 \quad -c_3] \tag{30}$$

$$\mathbf{K_m} = k_k \mathbf{L^T L} \tag{31}$$

$$\mathbf{h} = [-\mathbf{I} \quad \mathbf{I}] \tag{32}$$

$$\mathbf{H} = \mathbf{h^T h} \tag{33}$$

$$\mathbf{K_g} = \frac{k_k \delta l_k}{l_k + \delta l_k} \mathbf{H} \tag{34}$$

Where $k_k$ is the stiffness of the truss member ($k_k = E_k b_k w_k / l_k$ for a tape) and $\delta l_k$ the elongation of the deformed element. Its contribution in the internal force vector is:

$$\mathbf{Q_i} = k_k \delta l_k \mathbf{L^T} \tag{35}$$

The external force vector $\mathbf{Q_e}$ contains the components of the external load acting on the system. Once all contributions are assembled in the linear system, the mechanical equilibrium is obtained by updating the nodal displacement according to the following iterative scheme:

$$\mathbf{u} + (\mathbf{K_m} + \mathbf{K_g})^{-1}(\mathbf{Q_e} - \mathbf{Q_i}) \to \mathbf{u} \tag{36}$$

The 2-norm of the residual $\|\mathbf{Q_e} - \mathbf{Q_i}\|$ is used as convergence criterion.

In order to control the tape delaminations, the external load is incremented iteratively and the total potential energy variation is calculated at each increment. The delamination of a discrete length $\Delta l$ of a member in contact with the substrate leads to a modification of its length (and therefore its stiffness) and of the coordinates of the node in contact:

$$l + \Delta l \to l \tag{37}$$

$$\mathbf{x} + \Delta \mathbf{x} \to \mathbf{x} \tag{38}$$

The change in the attached node coordinate depends on the direction of the attached part of the delaminating tape respect to the detached one. It has two components on the x-y substrate plane and a zero value in the z direction ($\Delta l = \sqrt{\Delta x + \Delta y}$). At each step of the simulation, the energy variation is verified for all the members in contact, between the current state (a) and the state where the considered tape detachment has been incremented (b). The work variation associated with the detachment of the tape $k$ is the 1-norm of the product of the external force vector and the difference between displacement after and before detachment.

$$\frac{\Delta V}{\Delta l} = |\mathbf{Q}_{\text{ext}}(\mathbf{u_b} - \mathbf{u_a})| \tag{39}$$

The variation of elastic energy in the system is obtained as:

$$\frac{\Delta U_e}{\Delta l} = \frac{1}{2}\sum_{k=1}^{N}\left[\left(k_k \delta l_k^2\right)_b - \left(k_k \delta l_k^2\right)_a\right] \tag{40}$$

where $N$ is the total number of truss members in the system. Detachment occurs when:

$$\frac{\Delta V}{\Delta l} - \frac{\Delta U_e}{\Delta l} > wG \tag{41}$$

Tape fracture is introduced by a simple removal rule when one of the tape tensions reaches the critical value $T_c$. If the tape geometry and loading are in the same plane, the system can be reduced to a 2-D problem.

### 4. Results

We use the previously described numerical method to model the complete staple-pin system shown in *Figure 1*. The dragline, or cable that supports the external load is assumed to have a diameter $d^* = 0.2$ mm with a circular cross section $A^* = \pi d^{*2}$, an elastic modulus $E^* = 100$ MPa and a length $L^* = 100$ mm, for a total number of $N^* = 50$ transversal tapes. We assume for each of these tapes the following properties: $w = 1$ mm, $b = 0.01$ mm, $E = 100$ MPa, $l_a = 50$ mm, $\theta_0 = \pi/16$, $l_d = d^*/\sin\theta_0$ and $T_C = 0.2$ N. An example of the global load response, together with the evolution of the deformed and peeled system is shown in *Figure 6*.

We first consider the case in which the adhesive energy $G$ is small enough for the system to completely peel-off over its entire attached regions without any fracture ($G = 0.04$ MPa.mm.), applying a load perpendicular to the substrate at $\theta = \pi/2$. The corresponding overall load-displacement curve is shown in *Figure 6.A*, displaying an initial quasi-linear behaviour, and then a constant-load plastic branch, where an equilibrium is reached as the secondary transversal tapes are delaminating. The whole structure detaches at approximately constant load, apart from small oscillations in the force value due to the delaminating tapes. Various snapshots of the deformation profile of the entire structure as it delaminates are shown in *Figure 6.B*, highlighting the advancing delamination front as the load application point displacement increases. The overall adhesive force $F$ as a function of the adhesive energy is shown in *Figure 6.C*, again for an external load perpendicular to the substrate $\varphi = \pi/2$. This highlights the discussed transition from delamination to fracture behaviour of the tapes at $G_1 = 0.05$ MPa.mm. This value coincides with the optimal adhesive energy predicted with Eq. (20).

To analytically predict the global peeling force, energy balance expressed in Eq. (2) can be applied to the cable. The interaction between the cable and the substrate occurs through the transversal tapes, each of which can dissipate a total amount of energy $W$ given by Eq. (17). Dividing this value by the width of the tapes gives the energy per unit length needed to detach the cable. Using Eq. (11), the energy balance applied directly to the cable becomes:

$$F(1 - \cos\theta) + \frac{F^2}{2E^*A^*} = \frac{W}{w} \qquad (42)$$

Thus, the global peeling force becomes:

$$F = 2E^*A^*\left(\cos\theta - 1 + \sqrt{(1-\cos\theta)^2 + \frac{2W}{wE^*A^*}}\right) \qquad (43)$$

Comparing the maximum peeling force obtained in the simulations with the theoretical one (indicated as $F_1$) from Eq. (20), Eq. (21) and Eq. (43), a good agreement is seen (*Figure 6.C*).

Thus, the global adhesive force of the system is obtained from the energy dissipated by discrete sub-regions rather than from the maximum force they can carry. This has non-trivial consequences. In biological adhesion, typical structures often display a hierarchical architecture, which means that energy dissipation mechanisms occur at different scale levels simultaneously, each of them having a specific response specific load distributions over its sub-units. At present, most of the studies are focused on the detachment force of the contact units [23]. The present work shows that from a lower to and upper level, the dissipated energy of each contact is more important than its maximum detachment force. This is particularly important in cases in biological adhesion where most of the contacts are performed using tape-like units, displaying a typical "elastoplastic" behaviour such as that shown in *Figure 4*, where the maximum detachment force is not sufficient to determine the total dissipated energy.

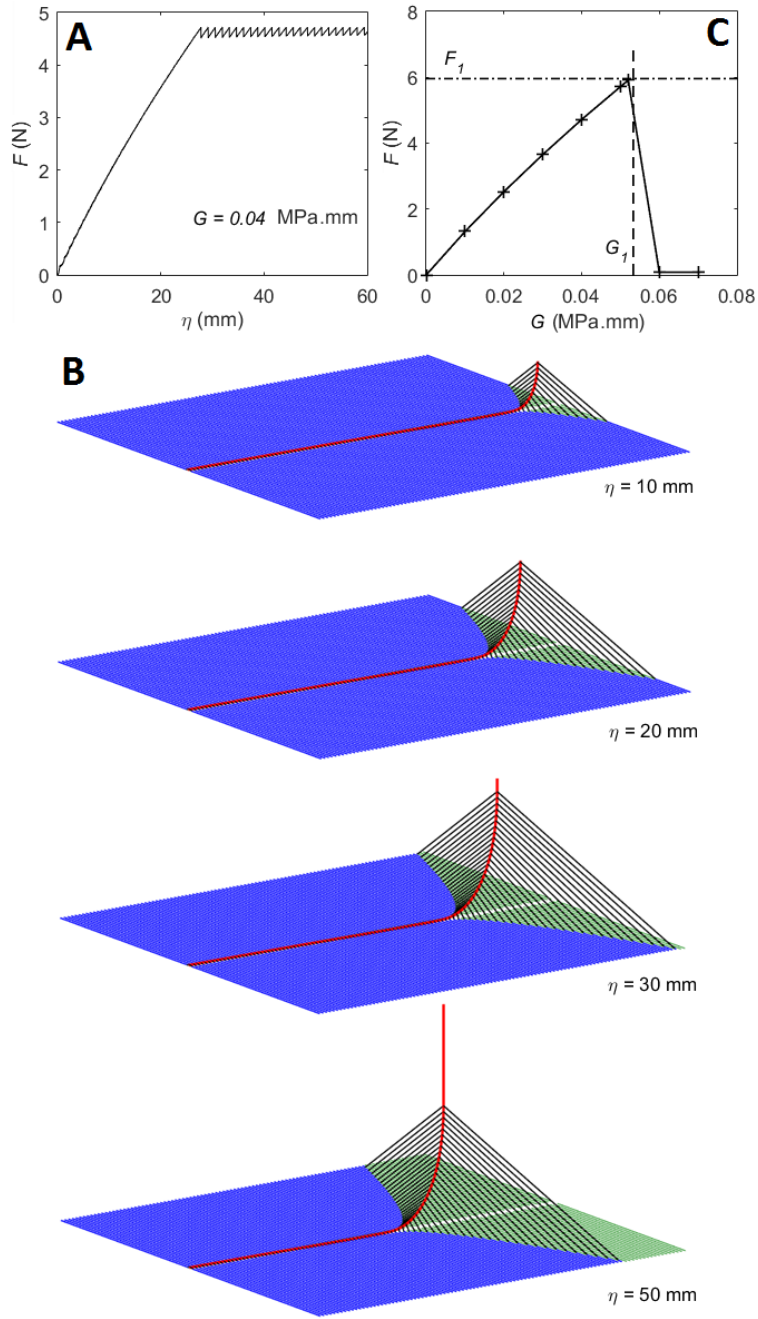

Figure 6: A. External load to displacement of the load application point during detachment of the staple-pin structure obtained from the numerical model. B. Corresponding evolution of the system (the red line is the cable, the black lines are the detached tape regions, the blue lines are the attached tape regions and the green line are the detachment path of the adhesive tapes. The fully

*detached tapes are not represented). C. Overall detachment force as a function of the adhesive energy obtained from the numerical model, plotted together with the theoretical optimal.*

## Conclusions

In conclusion, we have studied fibrous or tape-like attachment systems with multiple contacts, such as those found in staple-pin structures in spider webs, introducing a general analytical scheme that includes both delamination and tape fracture, and validating it with numerical simulations. We have shown that adhesive energy and mechanical strength are synergetic in providing optimized load-bearing properties, i.e. the maximum load an attachment can support before detachment. Additionally, we have shown that the energy dissipated by the contacts, accounting both for elastic deformation and detachment, determines the adhesive force of a multiple peeling system. Since structures formed by arrays of contact units, usually tape-like contacts, are recurrent in biological adhesives, the model discussed could help improve the understanding of Nature's strategies to enhance and optimize adhesion. This approach could also be useful in future for the design and optimization of artificial bioinspired adhesives.

## Acknowledgements

F.B. is supported by H2020 FET Proactive "Neurofibres" grant No. 732344, by the project "Metapp" (n. CSTO160004) funded by Fondazione San Paolo, and by the Italian Ministry of Education, University and Research (MIUR), under the "'Departments of Excellence'" grant L. 232/2016. This work was carried out within the COST Action CA15216 "European Network of Bioadhesion Expertise: Fundamental Knowledge to Inspire Advanced Bonding Technologies". N.M.P. is supported by the European Commission H2020 under the Graphene Flagship Core 2

No. 785219 (WP14 "Composites") and FET Proactive "Neurofibres" grant No. 732344 as well as by the Italian Ministry of Education, University and Research (MIUR), under the "'Departments of Excellence'" grant L. 232/2016. Computational resources were provided by the C3S centre at the University of Torino (c3s.unito.it ) and by hpc@polito (www.hpc.polito.it).